\documentclass[pra,floatfix,twocolumn]{revtex4}
\usepackage[T1]{fontenc}
\usepackage{amssymb,amsmath}
\usepackage{txfonts}
\usepackage{color,graphicx}

\begin{document}
\title[MD simulation analysis of DCCM in mOR-EG]
{
Molecular dynamics simulation analysis of 
structural dynamic cross correlation induced by odorant hydrogen-bonding in mouse eugenol olfactory receptor\footnote{
Biophys. Physicobiol. \textbf{21}, e210007 (2024).\\
\hspace*{0.4em}https://doi.org/10.2142/biophysico.bppb-v21.0007}
}
\author{Chisato Okamoto and Koji Ando\footnote{E-mail: ando\_k@lab.twcu.ac.jp}}

\affiliation{Department of Information and Sciences, 
    Tokyo Woman's Christian University, 
    2-6-1 Zenpukuji, Suginami-ku, Tokyo 167-8585, Japan}

\begin{abstract}
Structural fluctuations and dynamic cross-correlations
in the mouse eugenol olfactory receptor (Olfr73)
were studied by molecular dynamics (MD) simulation
to characterize the dynamic response of the protein upon ligand binding.
The initial structure was generated by the artificial intelligence tool AlfaFold2
due to the current lack of experimental data.
We focused on the hydrogen (H) bond of the odorant eugenol 
to Ser113, Asn207, and Tyr260 of the receptor protein, 
the importance of which has been suggested by previous experimental studies.
The H-bond was not observed in docking simulations, 
but in subsequent MD simulations the H-bond to Ser113 was formed in 2--4 ns.
The lifetime of the H-bond was in the range of 1--20 ns.
On the trajectory with the most stable (20 ns) H-bond, 
the structural fluctuation of the $\alpha$-carbon atoms of the
receptor main chain was studied by calculating the root mean square fluctuations,
the dynamic cross-correlation map, and the time-dependent dynamic cross-correlation.
The analysis suggested a correlation transfer pathway
Ser113 $\to$  Phe182 $\to$ (Leu259 or Tyr260) $\to$ Tyr291 
induced by the ligand binding with a time scale of 4--6 ns.
\end{abstract}

\maketitle

\section{Introduction}
The mammalian olfactory systems are capable of detecting and discriminating between
a wide range of chemically diverse molecules 
\cite{Buck1991,Kraft2000,Firestein2001,Zozulya2001,Zhang2002,Serizawa2003,Malnic2004,Reed2004,Sell2014}.
Olfactory receptors (ORs) are located on the cell membrane of olfactory nerves in the nasal cavity. 
Binding of odorant molecules to ORs leads to the dissociation of heterotrimeric G-proteins 
from the intracellular domain of the OR protein,
which transduces chemical signals into neuronal electrical responses.
The OR proteins belong to class A of G-protein coupled receptors (GPCRs) \cite{Yang2021},
which are characterized by seven transmembrane (TM) helices.
Due to the current lack of experimental 3D structural data of OR proteins,
molecular modeling studies of the OR mechanisms are limited 
\cite{DiPizio2014,Doszczak2007,Gelis2011,Floriano2000,Floriano2004,Katada2005,Baud2011,Baud2015,Yuan2019}.
Recent studies have mainly used homology modeling techniques
with the available 3D structures of other class A proteins 
such as bovine rhodopsin and human $\beta_2$-adrenergic receptor
as templates to construct 3D structural models of OR proteins.

\begin{figure}[h]
\centering
\includegraphics[width=0.4\textwidth]{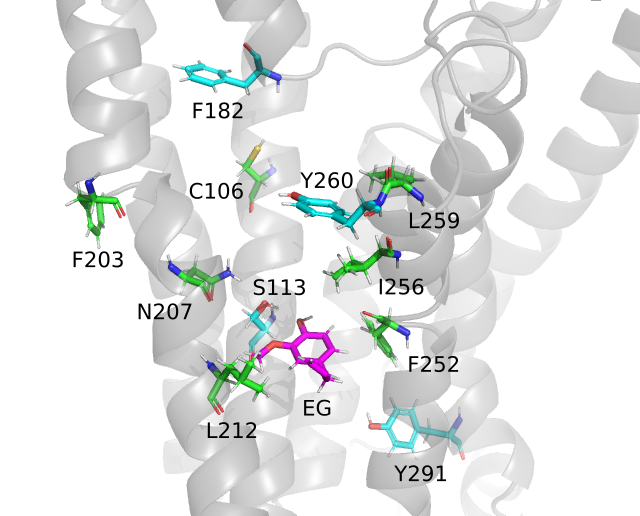}
\caption{A snapshot of the MD simulation showing the positions of the twelve key residues,
Cys106$^{3.33}$, 
Ser113$^{3.40}$, 
Phe182, 
Phe203$^{5.42}$, 
Asn207$^{5.46}$, 
Glu208$^{5.47}$, 
Leu212$^{5.51}$, 
Phe252$^{6.47}$, 
Ile256$^{6.51}$, 
Leu259$^{6.54}$, 
Tyr260$^{6.55}$,
and
Tyr291$^{7.53}$
(the superscripts denote Ballesteros-Weinstein numbers \cite{Ballesteros1995, PandySzekeres2022}),
relative to the odorant ligand eugenol (EG).
}
\label{fig:snapshotKeyRes}
\end{figure}

The mouse eugenol olfactory receptor Olfr73 (mOR-EG) has been studied as a prototype of mammalian ORs.
By combining computational molecular modeling with experimental techniques such as site-directed mutagenesis,
eleven critical amino acid residues 
(Cys106, Ser113, Phe182, Phe203, Asn207, Glu208, Leu212, Phe252, Ile256, Leu259, Tyr260,
see Figure \ref{fig:snapshotKeyRes}) 
located in TM3, TM5 and TM6 and the second extracellular loop
were identified as
essential for the ligand binding \cite{Katada2005,Baud2011}.
Molecular dynamics (MD) simulations of the apo-form OR protein
suggested 
an interesting mechanism for the transfer of structural change information 
to a residue (Tyr291) linked to the G-protein \cite{Baud2015}.
Another MD simulation study with a fingerprint interaction analysis
showed that the binding pocket of Olfr73 is smaller but more flexible
than those of non-olfactory GPCRs \cite{Yuan2019}.

In this work,
we further investigate 
the dynamic response of Olfr73 to ligand binding.
The artificial intelligence tool AlphaFold2 \cite{Jumper2021,Akdel2022,Yang2023}
is used to 
construct a 3D structural model of Olfr73.
The dynamics of both the apo and holo forms are studied 
by calculating the root mean square fluctuations (RMSF),
the dynamic cross-correlation map (DCCM) \cite{McCammon1984,Valente_2020,Parida_2020,Wang_2022,Dash_2022}, 
and the time-dependent dynamic cross-correlation (TDDCC)
from MD trajectories.
The dynamic structural response induced by the ligand binding suggests
pathways of correlation transfer and their corresponding time scales.

\section{Methods}
\label{sec:methods}
\subsection{Structure Model from AlphaFold2}
We used the ColabFold web server \cite{Mirdita2022} 
to construct a 3D structural model of Olfr73. 
(The access date was July 8, 2022.)
The default options of ColabFold (AlphaFold2 with MMseqs2) were used.
The amino-acid sequence was taken from UniProt entry Q920P2.

\subsection{MD Simulation of Apo Form}
\begin{figure}[h]
\centering
\includegraphics[width=0.4\textwidth]{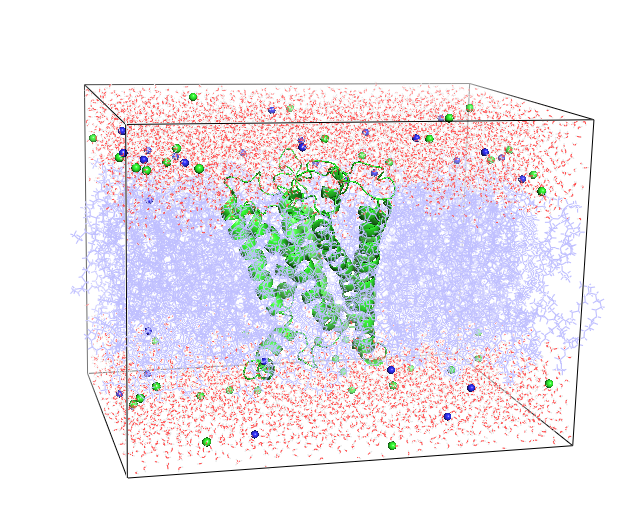}
\caption{A snapshot of the MD simulation box.}
\label{fig:snapshotBox}
\end{figure}

The structure model of Olfr73 from AlphaFold2 is embedded in a 
palmitoyl-oleolyl-glycelo-phosphatidyl-ethanolamine (POPE)
membrane in a saline solution,
using the membgene3 package of the myPresto5 software suite \cite{mypresto5}.  
The number of Na$^{+}$ and Cl$^{-}$ ions was adjusted to neutralize the system.
The resulting numbers of molecules and ions in the simulation box
were 229 POPE, 30 Na$^{+}$, 41 Cl$^{-}$, and 10815 H$_2$O.
After structural optimization by minimizing the total energy
with positional constraints on the main chain atoms by harmonic potentials,
constant pressure and temperature (NPT) simulations were performed to 
heat and equilibrate the system at 310 K and 1 atm.
The simulation time step was 0.5 fs.
The Berendsen method \cite{Berendsen1984} with a relaxation time of 100 fs was used 
to control the pressure and temperature.
Equilibration of the box lengths and the total potential energy was achieved 
in a 1.1 ns simulation.
The box lengths were then averaged over an additional 100 ps simulation.
The resulting box size was $105.9 \times 78.9 \times 77.7$ \AA$^3$ (Figure \ref{fig:snapshotBox}).
Next, a 20 ns NVT (constant volume and temperature) MD simulation was performed.
The Gaussian constraint method \cite{Evans1983,Nose1991} was used to control the temperature.
The simulation time step was 2 fs
with the SHAKE method \cite{Ryckaert1977,Kasahara2016} 
to constrain the bond lengths and angles involving hydrogen atoms.
The zero-dipole summation method \cite{Fukuda2011,Mashimo2013} 
was used to calculate the long-range electrostatic interaction.
We used the software psygene-G \cite{Mashimo2013} 
for the MD simulation.
The Amber99 \cite{Amber99} and Lipid14 \cite{Lipid14} potential force-fields were used.

\subsection{Structure Sampling and Odorant Docking}
The first 1 ns trajectory of the apo form was divided into twenty windows of 50 ps each.
One configuration per each window of the Olfr73 protein was randomly sampled. 
An eugenol molecule was docked 
to target points of the sampled configurations
near Ser113 and Asn207,
covering the spatial region surrounded by TM3, TM5 and TM6 helices.
The software sievgene \cite{sievgene} was used
for the docking simulation. 
The atomic charge parameters for eugenol were determined by the
electrostatic potential fitting with the RHF/6-31G(d) wave function.
the software GAMESS \cite{GAMESS} was used for the electronic structure calculations.

\subsection{MD Simulation of Holo Form}
The sampled and docked configurations of Olfr73 and eugenol were returned 
to the simulation box of the membrane solution.
The atomic coordinates of the membrane and the saline solution were optimized
by energy minimization
with the harmonic constraints to the positions of Olfr73 and eugenol.
After heating the temperature to 310 K in 100 ps,
the constant NVT MD simulation was performed for 10 ns.
In two out of twenty trajectories,
stable H-bonding between Ser113 and eugenol was observed for 2--3 ns.
The 3 ns H-bond trajectory was divided into ten 300 ps windows,
and one configuration per each window was randomly sampled.
For each configuration, 
the atomic velocities were randomized in the Maxwell-Boltzmann distribution
to restart the constant NVT MD simulation.
Six trajectories maintained the H-bond between eugenol and Ser113
for 1--20 ns.
From the 20 ns H-bond trajectory (Supplementary Figure S3),
the root mean square fluctuation (RMSF),
the dynamic cross-correlation map (DCCM) \cite{McCammon1984,Valente_2020,Parida_2020,Wang_2022,Dash_2022},
and the time-dependent dynamic cross-correlation (TDDCC)
were calculated.

\section{Results and Discussion}
\label{sec:results}
\subsection{Structure from AlphaFold2}
The resulting 3D structure from AlphaFold2
contained a disulfide bond between Cys98 and Cys180.
The structure was compared with 
the more recent UniProt entry (AF-Q920P2-F1-model\_v4.pdb)
in the Supplementary Figure S1.
The root mean square deviation of the aligned part (calculated with the PyMOL software \cite{pymol}) was 0.240 \AA.

\subsection{Sampling of Hydrogen-Bonded Trajectories}
\label{sec:sampleHBtraj}
\begin{figure}[h]
\centering
\includegraphics[width=0.4\textwidth]{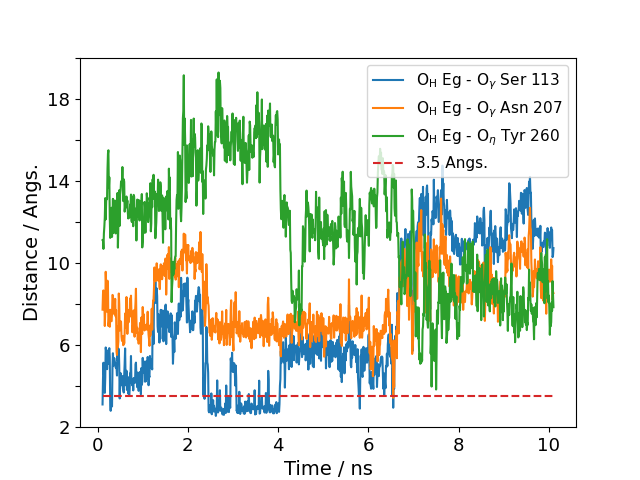}
\includegraphics[width=0.4\textwidth]{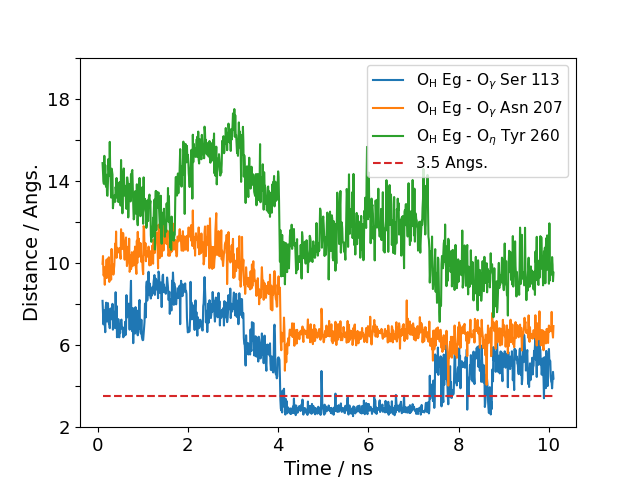}
\caption{Representative trajectories of the distance between the O$_\mathrm{H}$ atom of eugenol 
and the O atoms of Ser113, Asn207, and Tyr260, starting from the docking configurations.}
\label{fig:MDfromDock}
\end{figure}

Two representative trajectories 
of the distance between the O$_\mathrm{H}$ atom (the oxygen atom of the OH group) of eugenol 
and the O$_\gamma$ atom of Ser113,
the O$_\gamma$ atom of Asp207, and the O$_\eta$ atom of Tyr260
starting from the docking configurations 
are shown in Figure \ref{fig:MDfromDock}.
Since the O-O distance shorter than 3.5 \AA\ can be considered to form a H-bond,
both panels in Figure \ref{fig:MDfromDock} indicate that
a stable H-bond is formed between eugenol and Ser113 for 2--3 ns.
For Asp207, Glu208, and Tyr260,
stable H-bonds with eugenol were not observed 
in Figure \ref{fig:MDfromDock}
or in the other eighteen sampled trajectories
shown in the Supplementary Figure S2.

\begin{figure}[h]
\centering
\includegraphics[width=0.4\textwidth]{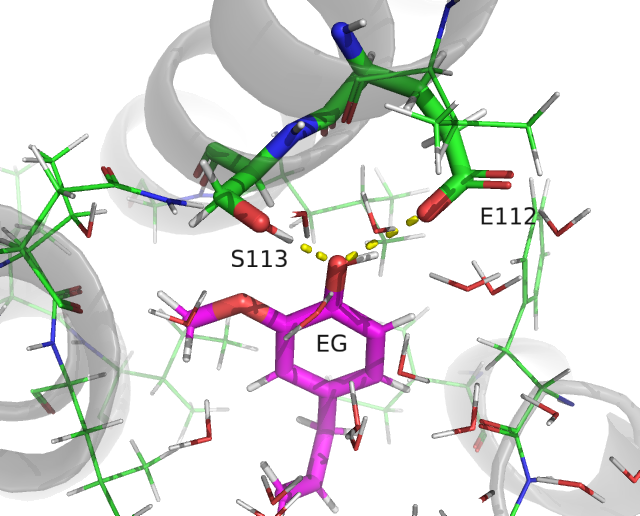}
\caption{A snapshot of the hydrogen-bond structure between eugenol (EG) and Ser113$^{3.40}$ and Glu112$^{3.39}$.}
\label{fig:snapshotHB}
\end{figure}

Figure \ref{fig:snapshotHB} shows a snapshot from the MD trajectory of the H-bond structure between eugenol and Ser113,
indicating that eugenol is also H-bonded to Glu112.
The double H-bond would be the main reason for the long lifetime ($> 2$ ns) of the ligand binding.

\subsection{Root-Mean-Squares Fluctuation}
\label{sec:rmsf}
Next, we examine the fluctuation of the protein structure induced by the ligand binding.
The root mean square fluctuation (RMSF) of the Cartesian coordinate $\boldsymbol{r}_i$ of the $i$-th atom
is defined by
\begin{equation}
\mathrm{RMSF}_{i}
= \sqrt{\langle \Delta \boldsymbol{r}_i^2 \rangle}
\end{equation}
where $\Delta \boldsymbol{r}_i =  \boldsymbol{r}_i - \langle \boldsymbol{r}_i \rangle$
is the deviation from the average $\langle \boldsymbol{r}_i \rangle$.
In this work, 
the fluctuations of the amino-acid residues are represented by those of the $\alpha$-carbon atoms. 

\begin{figure}[h]
\centering
\includegraphics[width=0.4\textwidth]{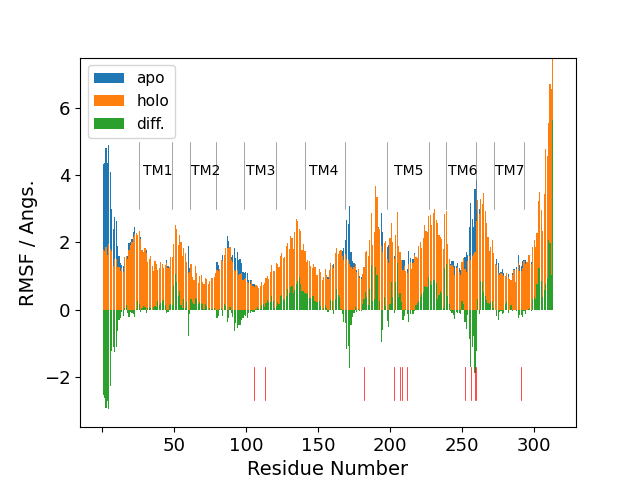}
\includegraphics[width=0.4\textwidth]{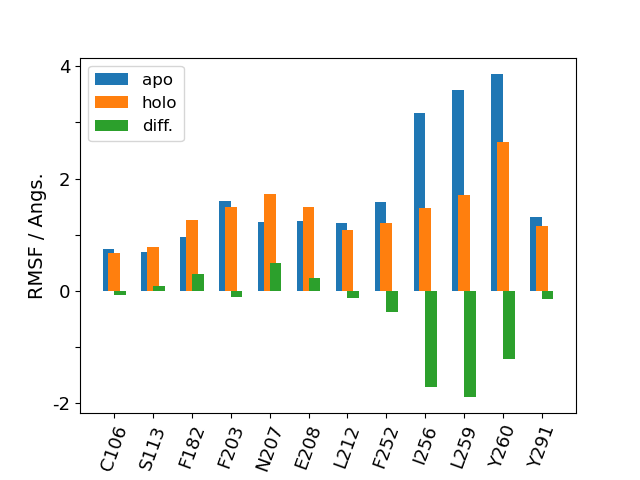}
\caption{Root-mean-square fluctuation (RMSF) of the C$_\alpha$ atoms of Olfr73 in the apo and holo forms 
    and their difference (holo minus apo). 
The left panel shows all residues. 
TM1--TM7 denote the transmembrane $\alpha$-helices.
The vertical lines below the bar plot indicate the twelve key residues shown in the right panel.}
\label{fig:rmsf}
\end{figure}

The calculated RMSFs in the apo and holo forms and their difference (holo minus apo) are shown in Figure \ref{fig:rmsf}.
In both apo and holo forms, 
the fluctuation is comparatively smaller in the central regions of the TM helices.
The RMSF is suppressed in the holo form for small residue numbers (1--12), in the beginning of TM3, 
after TM4, and in the latst region of TM6.
On the other hand, the RMSF is enhanced in the holo-form in the regions between TM1 and TM2, between TM3 and TM4,
before TM5, between TM5 and TM6, and after TM7.
T welve key residues are taken from the left panel of Figure \ref{fig:rmsf} and displayed in the right panel.
Eleven of them 
(Cys106, Ser113, Phe182, Phe203, Asn207, Glu208, Leu212, Phe252, Ile256, Leu259, and Tyr260)
are those reported in ref \cite{Baud2011} as essential for the olfactory function,
and one (Tyr291) is bonded to the G-protein \cite{Baud2015}.
A snapshot of their positions is shown in 
Figure \ref{fig:snapshotKeyRes}.
It can be seen 
in Figure \ref{fig:rmsf}
that the suppression of RMSF in the holo-form is remarkable for Ile256, Leu259, and Tyr260,
which are located near the end or TM6.

\begin{figure}[t]
\centering
\includegraphics[width=0.4\textwidth]{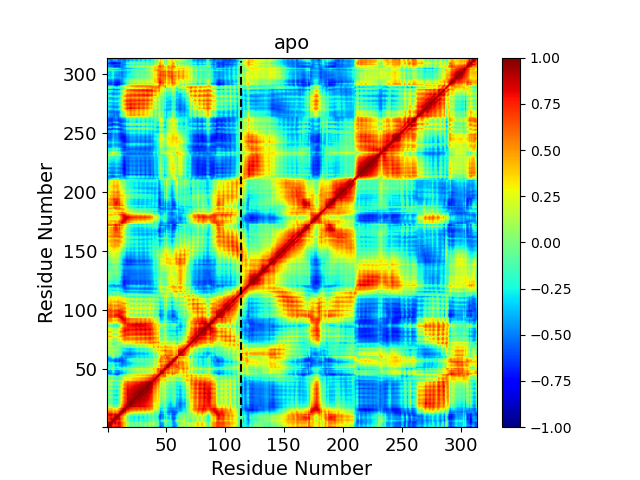}
\includegraphics[width=0.4\textwidth]{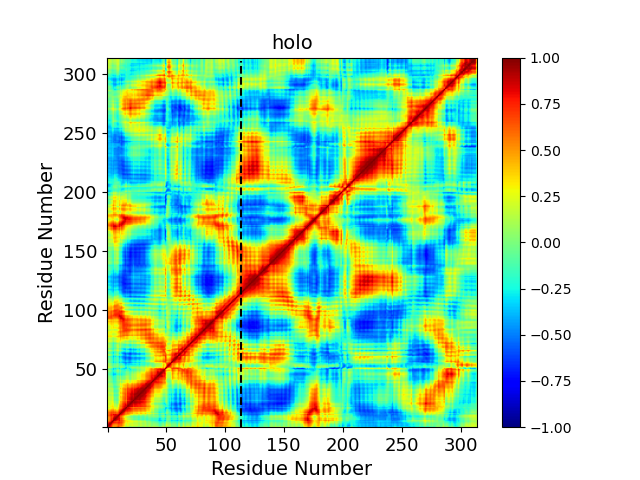}
\includegraphics[width=0.4\textwidth]{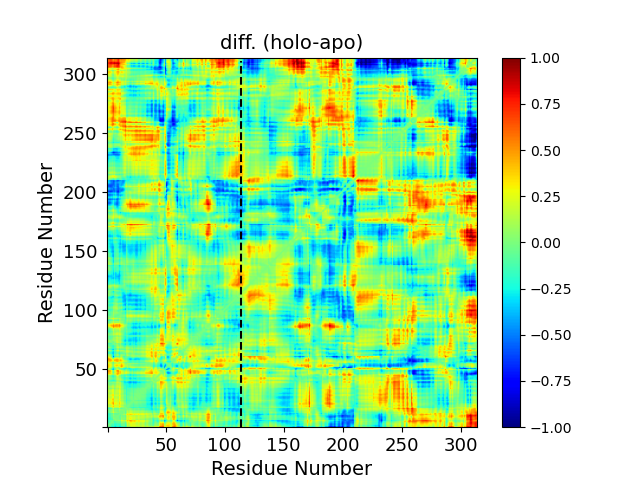}
\caption{Dynamic cross-correlation map (DCCM) of the C$_\alpha$ atoms of Olfr73 in the apo and holo forms 
and their difference (holo minus apo).}
\label{fig:dccm}
\end{figure}

\begin{figure}[h]
\centering
\includegraphics[width=0.4\textwidth]{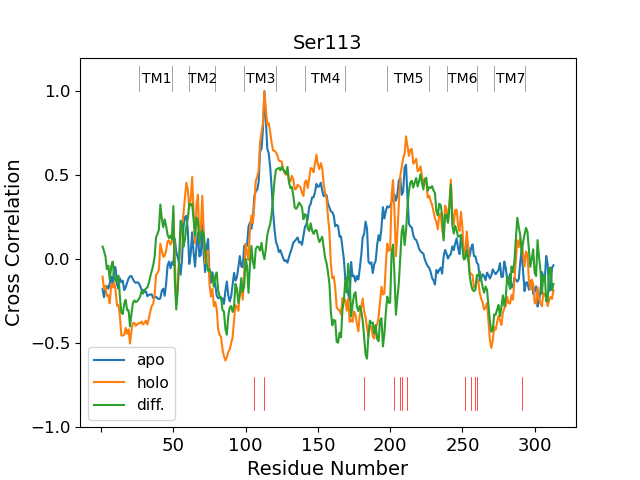}
\includegraphics[width=0.4\textwidth]{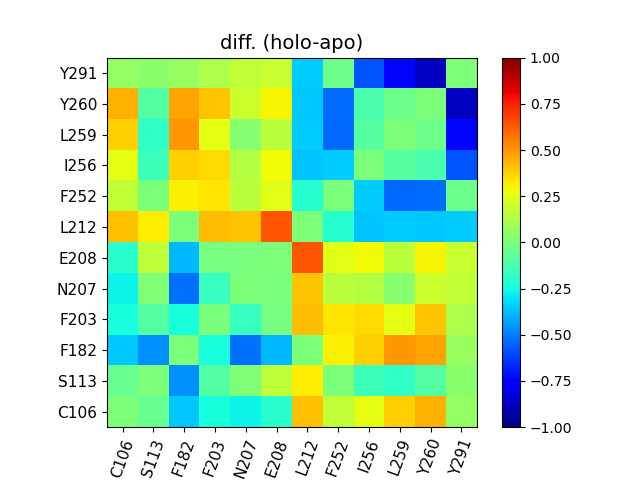}
\caption{Sections of the DCCMs in Figure \ref{fig:dccm} at Ser113 and 
difference of the DCCMs (holo minus apo) for the key residues.}
\label{fig:dccmDiff}
\end{figure}

\subsection{Dynamic Cross Correlation Map}
\label{sec:dccm}
The dynamic cross-correlation map (DCCM) \cite{McCammon1984,Valente_2020,Parida_2020,Wang_2022,Dash_2022}
is a map of the cross-correlation or covariance
between the fluctuations of the atomic Cartesian coordinates $\boldsymbol{r}_{i}$ and $\boldsymbol{r}_{j}$,
defined by
\begin{equation}
C_{ij}
= \frac
{\langle \Delta \boldsymbol{r}_i \cdot \Delta \boldsymbol{r}_j \rangle}
{\sqrt{\langle \Delta \boldsymbol{r}_i^2 \rangle} \sqrt{\langle \Delta \boldsymbol{r}_j^2 \rangle}} .
\label{eq:dccm}
\end{equation}
The numerator is the average of the inner product between the coordinate shift vectors.
The denominator is the normalization factor, which limits the value between $-1$ and $+1$.
The DCCM is close to $+1$ when the atomic coordinate fluctuations are in the same direction,
close to $-1$ when they are in the opposite direction, and close to $0$ when there is no directional correlation.
Analysis of the DCCM would reveal the
correlated dynamics of structural fluctuations,
and in particular, provide the key to identifying the correlation transfers 
from Ser113 H-bonded to the ligand to Tyr291 linked to the G-protein.
(See the lines below Eq. (\ref{eq:tddcc}) for the term \lq\lq correlation transfer\rq\rq).

The calculated DCCM for the apo and holo forms and their difference (holo minus apo) are shown
in Figure \ref{fig:dccm}.
Their cross sections at Ser113 are shown in the left panel of Figure \ref{fig:dccmDiff}.
(The corresponding plots at the other eleven key residues are displayed in Supplementary Figure S4.)
The difference of the correlations is remarkable;
between TM2 and TM3 (negative),
in the last part of TM3 (positive),
between TM4 and TM5 (negative),
in the last part of TM5 (positive),
between TM6 and TM7 (negative).
This means that negative or positive correlations in the structural fluctuation
are induced by the H-bond of the ligand eugenol to Ser113 and Glu112.

The difference map for the 12 key residues is shown in the right panel of Figure \ref{fig:dccmDiff}.
The map suggests several possible pathways of the correlation transfer.
In particular, we note the correlations between 
Ser113 and Phe182 (negative),
Phe182 and Leu259 (positive),
and Leu259 and Tyr291 (negative).
Leu259 and Tyr260 show similar behavior.
Therefore, one of the simplest correlation transfer pathways would be
Ser113 $\to$ Phe182 $\to$ (Leu259 or Tyr260) $\to$ Tyr291.

\begin{figure}[h]
\centering
\includegraphics[width=0.4\textwidth]{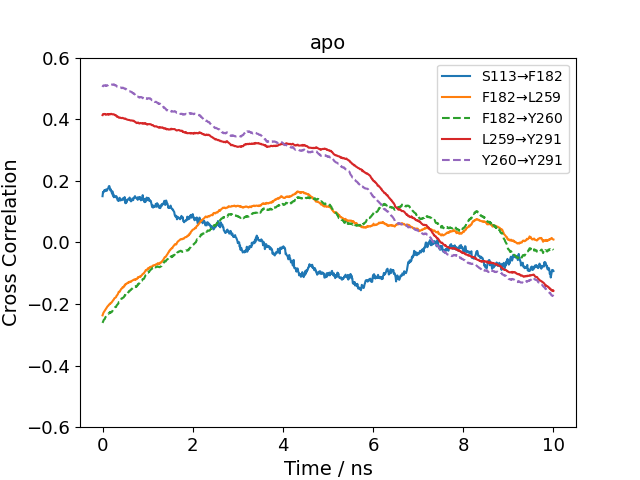}
\includegraphics[width=0.4\textwidth]{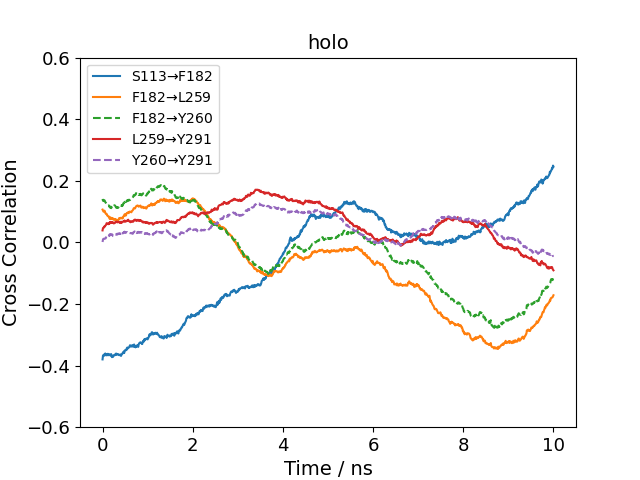}
\caption{Time-dependent dynamic cross-correlation (TDDCC) of the apo and holo forms
    of Olfr73 between the key residues
    identified as being in the major correlation transfer pathway.}
\label{fig:tddcc}
\end{figure}

\subsection{Time-Dependent Dynamic Cross Correlation}
\label{sec:tddcc}
For the correlation transfer pathway identified in the previous section,
we calculated the time-dependent dynamic cross-correlation (TDDCC) defined by
\begin{equation}
C_{ij}(t) 
= \frac
{\langle \Delta \boldsymbol{r}_i(0) \cdot \Delta \boldsymbol{r}_j(t) \rangle}
{\sqrt{\langle \Delta \boldsymbol{r}_i^2\rangle} \sqrt{\langle \Delta \boldsymbol{r}_j^2 \rangle}}
\label{eq:tddcc}
\end{equation}
This is the cross-correlation with the time delay $t$ of the fluctuations of the atomic coordinates
$\boldsymbol{r}_{i}$ and $\boldsymbol{r}_{j}$.
The DCCM of Eq. (\ref{eq:dccm}) is the simultaneous ($t=0$) part of the TDDCC.
Thus, the TDDCC is a time-dependent extension of the DCCM.
(To our knowledge, it has never been defined and computed in the literature).
The amplitude of the DCCM measures the correlation of the structural fluctuations,
and its sign indicates the parallel or antiparallel correlation.
Therefore, the time evolution of the TDDCC would be informative on the time scale of the correlation transfer.
(More precisely, in this paper we call the behavior of the TDDCC 
the \lq\lq correlation transfer\rq\rq).
Since we compute the TDDCC from the equilibrium simulation,
we assume that 
the essential dynamics of the non-equilibrium response 
induced by the ligand binding
is embedded in the time correlation in the equilibrium 
(fluctuation-dissipation theorem).

The calculated TDDCC along the path
Ser113 $\to$ Phe182 $\to$ (Leu259 or Tyr260) $\to$ Tyr291
is shown in Figure \ref{fig:tddcc}.
The results with Leu259 and Tyr260 are almost identical.
All TDDCCs change their qualitative behavior after the ligand binding.
The TDDCC between Ser113 and Phe182 is overall a decaying function of time in the apo form,
while it is basically an increasing function of time in the holo form.
Interestingly, they have minimum or maximum at $t \simeq 6$ ns.
The TDDCC between Phe182 and Leu259 (or Tyr260) decays from positive 
to negative in 10 ns in the apo form,
while in the holo form it oscillates near zero on an ns timescale.
The TDDCC between Leu259 (or Tyr260) and Tyr291 increases from negative
to positive in $\sim 4$ ns and stays near zero (or slowly decays to zero) in the apo form,
while it decreases from positive to negative overall with oscillation
in ns timescale.

\section{Conclusion}
\label{sec:concl}
Structural fluctuations and dynamic cross-correlations
in the mouse eugenol olfactory receptor Olfr73
were studied by molecular dynamics (MD) simulation.
Among the candidate residues for H-bonding with the odorant ligand eugenol, Ser113, Asn207, and Tyr260,
we find that only Ser113 forms a stable H-bond in the sampled trajectories.
The lifetime of the H-bond was in the range of 1--20 ns.
The structural fluctuations of the C$_\alpha$ atoms of the
receptor main chain were investigated by calculating the root mean square fluctuations (RMSF),
the dynamic cross-correlation map (DCCM), and the time-dependent dynamic cross-correlation (TDDCC).
The analysis suggested a correlation transfer pathway
Ser113 $\to$  Phe182 $\to$ (Leu259 or Tyr260) $\to$ Tyr291 
induced by the ligand binding.
The TDDCC indicated that the time scale of the correlation transfer was 4--6 ns.
However, to our knowledge, there is no experimental evidence to verify these results.
Time-dependent spectroscopies such as fluorescence resonance excitation transfer (FRET) \cite{Nguyen2022}, 
resonance Raman \cite{Yamashita2022}, and transient-grating \cite{Terazima2021}
could be useful to clarify the time scale of the structural correlation transfer.

In this work, we focused on the twelve key residues identified in the previous experimental studies.
The correlation transfer pathways highlighted in this work appear to be the most straightforward ones
that directly involve the key residues.
More extensive analysis would be required to clarify the role of other residues 
that may be directly or indirectly involved in the correlation transfer.
In addition, the precise mechanism by which the dissociation of the G-protein is induced is still unclear.
Work on these issues is ongoing.





\section*{Acknowledgments}
We thank Dr. S. Katada and Prof. K. Touhara for useful discussions.
This work was supported by JSPS KAKENHI Grant Number JP19K22173.

\end{document}